\documentclass{iaus}
\usepackage{graphicx}
\def\msun{{\rm M_{\odot}}}

\def\me{{\dot M_{\rm Edd}}}
\def\md{{\dot M_{\rm dyn}}}
\def\mo{{\dot M_{\rm out}}}
\def\le{{L_{\rm Edd}}}
\def\rsh{{R_{\rm shock}}}
\def\rinf{{R_{\rm inf}}}

\title[JD 11.~~Accretion and Outflow in AGN] %% give here short title %%
{Accretion and Outflow in Active Galaxies}

\author[Andrew King]   %% give here short author list %%
{Andrew King$^1$}

\affiliation{$^1$Theoretical Astrophysics Group, \\ University of
  Leicester\\ Leicester LE1 7RH, U.K. \\ email: {\tt
    ark@astro.le.ac.uk} \\[\affilskip]}

\pubyear{2009}
\volume{267}  %% insert here IAU Symposium No.
\pagerange{119--126}
\jname{Co-Evolution of Central Black Holes and Galaxies}
\editors{B.M.\ Peterson, R.S.\ Somerville, \& T.\ Storchi-Bergmann, eds.}
\begin{document}

\maketitle

\begin{abstract}

I review accretion and outflow in active galactic nuclei.  Accretion
appears to occur in a series of very small--scale, chaotic events,
whose gas flows have no correlation with the large--scale structure of
the galaxy or with each other. The accreting gas has extremely low
specific angular momentum and probably represents only a small
fraction of the gas involved in a galaxy merger, which may be the
underlying driver.

Eddington accretion episodes in AGN must be common in order for the
supermassive black holes to grow. I show that they produce winds with
velocities $v \sim 0.1c$ and ionization parameters implying the
presence of resonance lines of helium-- and hydrogenlike iron. The
wind creates a strong cooling shock as it interacts with the
interstellar medium of the host galaxy, and this cooling region may be
observable in an inverse Compton continuum and lower--excitation
emission lines associated with lower velocities.  The shell of matter
swept up by the shocked wind stalls unless the black hole mass has
reached the value $M_{\sigma}$ implied by the $M - \sigma$
relation. Once this mass is reached, further black hole growth is
prevented. If the shocked gas did not cool as asserted above, the
resulting (`energy--driven') outflow would imply a far smaller SMBH
mass than actually observed. Minor accretion events with small gas
fractions can produce galaxy--wide outflows, including fossil outflows
in galaxies where there is little current AGN activity.


\end{abstract}

\firstsection % if your document starts with a section,

\section{Accretion: large--scale}

Accretion on to a black hole is the most efficient way of extracting
energy from normal matter, and so must power the most luminous
phenomena in the Universe. To drive quasars and other bright AGN
without exceeding the Eddington limit requires supermassive black hole
(SMBH) accretors, ranging up to several $10^9\msun$, and accretion rates of
up to $10\msun~{\rm yr}^{-1}$. These statements probably encapsulate
all that is securely known about accretion in AGN.

The $M - \sigma$ relation (see below) strongly suggests a connection
between SMBH and galaxy growth. This in turn points to galaxy mergers
as the common motor of both phenomena. Cosmological simulations
(e.g. Di Matteo et al., 2005) aim to show the plausibility of this
idea, by demonstrating how a series of mergers can produce SMBH and
galaxies satisfying the $M- \sigma$ relation at low redshift. To make
the calculations tractable requires a sub--resolution recipe
for accretion, and this is usually taken as the Bondi rate
\begin{equation}
\dot M_B = 4\pi R_B^2\rho c_s,
\label{bondi}
\end{equation}
where
\begin{equation}
R_B = {2GM\over c_s^2}
\label{rbondi}
\end{equation}
is the Bondi radius, with $c_s$ the local sound speed, $\rho$ the
gas density and $M$ the SMBH mass.

However there are several problems with this recipe. First, it is
self--consistent only if $M$ is a good approximation to the total
gravitating mass inside $R_B$. This requires
\begin{equation}
R_B < {GM\over 2f_g \sigma^2} \sim 10 - 20~{\rm pc},
\label{bondi2}
\end{equation}
where $f_g \simeq 0.16$ is the gas fraction relative to dark matter,
and $\sigma$ is the velocity dispersion in the galaxy bulge. This is
far smaller than the spatial resolution available in typical
cosmological simulations. If the resolution scale is $R > R_B$, the
recipe gives $\dot M \sim (R/R_B)^2\dot M_B >> \dot M_B$. 
%If one takes
%$c_s \sim \sigma$, as might be expected for consistency, the estimated
%$\dot M$ is close to the dynamical rate  $\dot M_{\rm dyn}$ discussed
%below (in eq \ref{dyn}). 
This often leads to estimated accretion rates far above the Eddington
rate $\me$, which have to be corrected by assuming that the rate never
goes above this value. Although this may be roughly correct for bright
quasars (see below), it is obvious that this arbitrary procedure must
give an entirely misleading impression of the duty cycle of accretion.

In fact it is unlikely that any AGN accretes at very super--Eddington
rates. For the maximum possible accretion rate is the dynamical value
\begin{equation}
\dot M_{\rm dyn} \simeq {f_g \sigma^3\over 2G},
\label{dyn}
\end{equation}
which describes the case where gas
is initially in rough virial equilibrium in the bulge of a galaxy with
velocity dispersion $\sigma$ and baryonic mass fraction
$f_g$. Parametrizing, we find
\begin{equation}
\md \simeq 1.4\times 10^2 \sigma_{200}^3~\msun\, {\rm yr}^{-1}
\label{dyn2}
\end{equation}
where $\sigma_{200} = \sigma/(200~{\rm km\, s^{-1}})$, and we have
taken $f_g = 0.16$. For a black hole mass close to the observed $M -
\sigma$ relation this implies an Eddington ratio 
\begin{equation}
\dot m < {\md \over \me} \simeq {33\over \sigma_{200}} \simeq
     {39\over M_8^{1/4}}
\label{eddrat}
\end{equation}
where $M_8 = M/10^8\msun$. Since $0.1 < M_8 < 10$ for the black holes
in AGN, and $\md$ is an upper limit to $\dot M$, modest values $\dot m
\sim 1$ of the Eddington ratio are likely. Indeed, in the case where
the SMBH does not dominate the mass inside the estimated Bondi radius,
a realistic estimate of the Bondi rate is actually close to the
dynamical value, since
\begin{equation}
\dot M_B = 4\pi R_B^2\rho c_s = 3{M_gc_s\over R_B}
\end{equation}
where $M_g = 4\pi R_B^3\rho/3$. Now using (\ref{rbondi}) with $M_g$ in
place of $M$ we see that
\begin{equation}
\dot M_B = {3\over 2}{c_s^3\over G}
\end{equation}
and in a realistic situation we would expect $c_s \sim \sigma$.

Even if one were able to resolve the Bondi radius and estimate the
rate $\dot M_B$ cleanly, it is still unlikely that this gives an
estimate of the true accretion rate at the black hole and thus the AGN
luminosity. The reason is that in any conceivable physical situation
the gas must have sufficient angular momentum to orbit the black hole,
and so must form an accretion disc. Thus the term `Bondi accretion' is
better rendered as `Bondi capture'. Gas inside the Bondi radius cannot
easily escape the black hole's vicinity, but is not required to
accrete on to it at the Bondi rate. 

\section{Accretion: discs}

The fact that accretion must ultimately proceed via a disc now leads
to another set of difficulties. If, as is likely in many cases, the
disc cools efficiently, it will become thin and Keplerian. In this
case we can compute its viscous timescale
\begin{equation}
{R^2\over \nu} = {2\times 10^{10}\over \alpha_{0.1}}\left({R\over
  10^3H}\right)^2{R_{\rm pc}^{3/2}\over M_8^{1/2}}~{\rm yr}
\label{visc}
\end{equation}
at disc radius $R = R_{\rm pc}~{\rm pc}$, where $\alpha_{0.1}$ is the
  viscosity parameter $\alpha$ in units of its likely value 0.1 (King
  et al., 2007), $H \simeq 10^{-3}R$ is the disc scaleheight, and $M_8
  = M/10^8\msun$.  So unless the disc is very small, its viscous time
  is too long for significant accretion on to the SMBH. If the disc is
  large, on the other hand, it is likely to become self--gravitating
  and fragment into stars, since its mass $M_{\rm disc}$ exceeds the
  self--gravity limit $\sim (H/R)M$, i.e.
\begin{equation}
{M_{\rm disc}\over M}{R\over H} = {0.2\over \alpha_{0.1}}\left({R\over
  10^3H}\right)^{3}\left({R_{\rm pc}\over M_8}\right)^{3/2}{L\over \le}
\label{frag}
\end{equation}

Equation (\ref{visc}) shows that gas orbiting at only a few parsecs
takes more than a Hubble time to accrete. So the gas which forms the
disc and ultimately accretes must have arrived very close to the SMBH,
with very little angular momentum. On would not expect such an
accurate aim for most of the gas involved in a galaxy merger, so most
of this gas evidently cannot accrete on to the SMBH. This is
reasonable, given that we know (H\"aring \& Rix 2004) that the mass of
the black hole is only about $10^{-3}$ of the galaxy bulge's baryonic
mass. The merger process is evidently extremely inefficient in feeding
the black hole, with most growing the bulge or other parts of the host
galaxy. 

Since the accretion disc is so small compared with the galaxy, and
involves so little of the gas involved in a merger, this also suggests
that its net angular momentum is likely to be uncorrelated with the
large--scale structure of the host. Confirmation of this comes from
the observed directions of AGN jets. As the jets are relativistic,
they must be launched from the very near vicinity of the black hole,
normal to the plane of the disc there. Their directions are observed
to be uncorrelated with the galaxy structure (Kinney et a., 2000), just
as the argument above suggests. Moreover, successive feeding events
seem to produce jets whose directions deviate significantly from the
previous ones.

We can now see an emerging picture of AGN accretion as a series of
very small--scale, chaotic events, whose gas flows have no correlation
with the large--scale structure of the galaxy or with each other. The
feeding events must have extremely low specific angular momentum
compared with that typical of the gas in a galaxy merger (King \&
Pringle, 2006, 2007). 

This picture does seem to work well in explaining some key features,
in particular the evolution of mass and spin in SMBH, and the jet
directions discussed above (King \& Pringle, 2006, 2007; King et al.,
2008; Fanidakis et al., 2009). The key here is that the black hole
spin specifies the efficiency $\eta$ of luminous energy release by
accretion, and thus the accretion luminosity 
\begin{equation}
L_{\rm acc} = \eta M c^2
\label{lacc}
\end{equation}
The higher the spin, the higher $\eta$, and thus the {\it
  lower} the rate at which the black hole mass can grow, because the
accretion luminosity cannot greatly exceed the Eddington limit. Hence
rapid black hole growth to high masses, as observed in some
high--redshift quasars (Barth et al., 2003; Willott et al., 2003),
requires low black--hole spin. However for some time attempts to
understand this process were frustrated because it was thought that
the Lense--Thirring effect would always quickly co--align a misaligned
accretion disc with the black hole spin (Scheuer \& Feiler, 1996). In
this case virtually all accretion takes place through a prograde disc,
leading to rapid spin--up to high values of the Kerr $a$
parameter. This made it impossible to understand the high SMBH masses
referred to above without appealing to initial black hole `seeds'
which were themselves already more massive than many SMBH in the
low--redshift Universe (cf Volonteri et al., 2005). The resolution of
this problem was the realization (King et al., 2005) that the
condition for co-- or counter--alignment actually depends on the
magnitudes of the disc and black hole angular momenta, and their
initial orientation. Scheuer \& Feiler's (1996) paper had implicitly
assumed conditions allowing only co--alignment and spinup. Using the
analytic formula of King et al., (2005) and assuming sufficiently
small feeding events (e.g. limited by self--gravity) shows that most
SMBH are likely to have low spins. The exception is a group in giant
ellipticals where a direct coalescence of two SMBH has produced a
rapid spin (King et al., 2008; Fanidakis et al., 2009) which
subsequent randomized gas accretion is too insignificant to dilute.

It appears that this general picture of small--scale, chaotic
accretion events is in reasonable accord with observations of
AGN. However reproducing these conditions theoretically is a challenge
for models of the feeding process (cf Hopkins \& Quataert,
2009). Similarly, the mechanics of the accretion disc itself,
particularly its innermost parts, is the subject of intense
research. At a fundamental level, it is now almost universally agreed
that magnetic fields are implicated in the `viscous' process removing
angular momentum from disc material and causing it to accrete (Balbus
\& Hawley, 1991). However numerical implementations of this idea are
not yet definitive (cf King et al., 2007). Simulations using the
shearing--box approximation appear to suggest that angular momentum
removal becomes less efficient as numerical resolution is increased
(Fromang \& Papaloizou, 2007), and as yet no simulation appears to
give viscosity as large as that deduced from observation without
making the assumption of a net vertical magnetic field (King et al.,
2007). Given these theoretical problems in describing disc accretion,
we are still some distance from a deterministic picture of it.

\section{Outflows}

All galaxies are likely to go through active phases as they grow by
mergers. Given the rarity of active galaxies among all galaxies, these
phases must be relatively short. Accordingly, AGN must feed at fairly
high rates to grow the observed high SMBH masses. There is no obvious
reason why these rates should respect the black hole's Eddington
limit, so outflows driven by continuum radiation pressure are a
natural consequence. This is an encouraging deduction, as outflows
driven by black holes offer a simple way of establishing relations
between the SMBH and its host galaxy, and hence potential explanations
for the $M - \sigma$ and $M - M_{\rm bulge}$ relations (Ferrarese \&
Merritt, 2000; Gebhardt et al. 2000; H\"aring \& Rix 2004).

However it is clear from (\ref{eddrat}) that the Eddington ratio $\dot
m$ is limited to modest values in AGN. (This contrasts strongly with
accretion in stellar--mass binary systems because their very short
dynamical timescales ($\sim$ orbital period) allow extremely high
dynamical mass transfer rates and hence $\dot m >> 1$ -- for example
the well--known binary SS433 has $\dot m \sim 5000$, cf King et al.,
2000; Begelman et al., 2006.) The electron scattering optical depth
$\tau$ in a quasi--spherical super--Eddington wind scales linearly
with $\dot m$, and is of order unity for $\dot m \sim 1$. This low
scattering depth implies that the total momentum of a $\dot m \sim 1$
AGN wind must be of order the photon momentum (King \& Pounds, 2003)
i.e.
\begin{equation}
\mo v \simeq {\le\over c},
\label{mom}
\end{equation}
as is for example also found for the winds of hot stars. Using
(\ref{lacc}) with $L_{\rm acc} = \le$ in (\ref{mom}) gives the wind
velocity
\begin{equation}
v \simeq {\eta\over \dot m}c \sim 0.1c.
\label{v}
\end{equation}

Since the wind moves with speed $\sim 0.1c$, it can persist long after
the AGN is observed to have become sub--Eddington. The duration of the
lag is $\sim 10R/c$, where $R$ is the radial extent of the wind. For
$R > 3$~pc this lag is at least a century, and far longer lags are
possible, as we shall see. This may be the reason why AGN showing
other signs of super--Eddington phenomena (e.g. narrow--line Seyfert 2
galaxies) are nevertheless seen to have sub--Eddington luminosities
(e.g. NGC 4051: Denney et al., 2009).

With (\ref{v}), the mass conservation equation for the outflow gives
the combination $NR^2 = \mo/4\pi v$ specifying the ionization parameter
\begin{equation} 
\xi = {L_i\over NR^2}
\label{ion}
\end{equation}
of the wind. Here $L_i = l_i\le$ is the ionizing luminosity, with
$l_i< 1$ a dimensionless parameter specified by the quasar spectrum,
and $N = \rho/\mu m_p$ is the number density. This gives
\begin{equation}
\xi = 3\times 10^4\eta_{0.1}^2l_2\dot m^{-2},
\label{ion2}
\end{equation}
where $l_2 = l_i/10^{-2}$, and $\eta_{0.1} = \eta/0.1$. 

Equation (\ref{ion2}) shows that the wind momentum and mass rates
determine its ionization parameter: for a given quasar spectrum, the
predominant ionization state is such that the threshold photon energy
defining $L_i$, and the corresponding ionization parameter $\xi$,
together satisfy (\ref{ion2}). This requires high excitation: a low
threshold photon energy (say in the infrared) would imply a large
value of $l_2$, but the high value of $\xi$ then given by (\ref{ion2})
would require the presence of very highly ionized species, physically
incompatible with such low excitation. For a typical quasar
spectrum, an obvious self--consistent solution of (\ref{ion2}) is $l_2
\simeq 1$, $\dot m \simeq 1$, $\xi \simeq 3\times 10^4$.  
This corresponds to a photon energy threshold appropriate for helium--
or hydrogenlike iron (i.e. $h\nu_{\rm threshold}\sim 9$~keV).

we conclude that

{\it Eddington winds from AGN are likely to have velocities $\sim
  0.1c$, and show the presence of helium-- or hydrogenlike iron.}

A number of such winds are known (see Cappi, 2006, for a review). This
Section shows that it is no coincidence that in all cases the wind
velocity is $v \sim 0.1c$, and further that they are all found by
identifying blueshifted resonance lines of Fe XXV, XXVI in
absorption. Conversely, any observed wind with these properties
automatically satisfies the momentum and mass relations, strongly
suggesting launching by an AGN accreting at a slightly
super--Eddington rate.

\section{Interaction with the host}

It is clear that an Eddington wind of the type discussed above can
have a significant effect on its host galaxy. The kinetic power of
the wind is 
\begin{equation}
\mo {v^2\over 2} = {v\over 2c}\le \simeq 0.05\le
\label{power}
\end{equation}
where we have used (\ref{mom}) and (\ref{v}). If the wind persists as
the hole doubles its mass (i.e. for a Salpeter time), its total energy
is $\simeq 5\times 10^{59}M_8$~erg, where $M_8$ is the hole mass in
units of $10^8\msun$. This formally exceeds the binding energy $\sim
M_{\rm bulge}\sigma^2 \sim 3\times 10^{58}$~erg of a galaxy bulge with
baryonic mass $M_{\rm bulge} \sim 10^{11}\msun$ and velocity
dispersion $\sigma \sim 200~{\rm km\, s^{-1}}$ (as suggested by the $M
- M_{\rm bulge}$ and $M - \sigma$ relations). Evidently the coupling
of wind energy to the galaxy must be inefficient, as black holes would
destroy or at least severely modify their host bulges in any
significant super--Eddington growth phase.  As in the corresponding
problem for a stellar wind, the interaction with the host must
successively involve an inner (reverse) shock, slowing the central
wind, a contact discontinuity between the shocked wind and the
shocked, swept--up interstellar medium, and an outer (forward) shock
driven into this medium and sweeping it outwards, ahead of the shocked
wind (see Fig. 1).

\begin{figure}[b]
% \vspace*{-2.0 cm}
\begin{center}
 \includegraphics[width=3.4in]{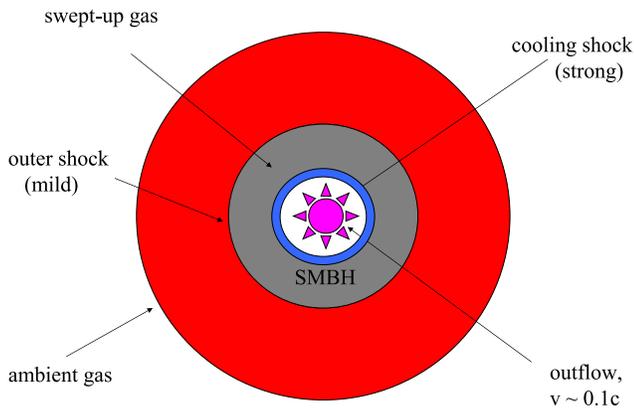} 
% \vspace*{-1.0 cm}
% \caption{Path of pre-solar grains from their stellar sources to the
%   laboratory.}
\caption{Schematic view of the shock pattern resulting from the impact
  of an Eddington wind on the interstellar gas of the host galaxy. A
  supermassive black hole (SMBH) accreting at just above the Eddington
  rate drives a fast wind (velocity $u = v \sim \eta c \sim 0.1c$),
  whose ionization state makes it observable in X--ray absorption
  lines. The outflow collides with the ambient gas in the host galaxy
  and is slowed in a strong shock. The inverse Compton effect from the
  quasar's radiation field rapidly cools the shocked gas, removing its
  thermal energy and strongly compressing and slowing it over a very
  short radial extent. This gas may be observable in an inverse
  Compton continuum and lower--excitation emission lines associated
  with lower velocities. The cooled gas exerts the preshock ram
  pressure on the galaxy's interstellar gas and sweeps it up into a
  thick shell (`snowplough'). This shell's motion drives a milder
  outward shock into the ambient interstellar medium. This shock
  ultimately stalls unless the SMBH mass has reached the value
  $M{_\sigma}$ satisfying the $M - \sigma$ relation.}

   \label{fig1}
\end{center}
\end{figure}

%\begin{figure*}

%\centerline{\psfig{file=diag.eps,width=0.5\textwidth,angle=0}}
%{epsfxsize3cm \epsfbox{diag.eps}
The inefficient coupling of wind energy to the galactic baryons noted
above strongly suggests that the shocked wind cools rapidly after
passing through the inner shock. This removes the thermal pressure
generated in the shock, and leaves only the preshock ram pressure
acting on the interstellar medium.

\begin{figure}[b]
% \vspace*{-2.0 cm}
\begin{center}
 \includegraphics[width=3.4in]{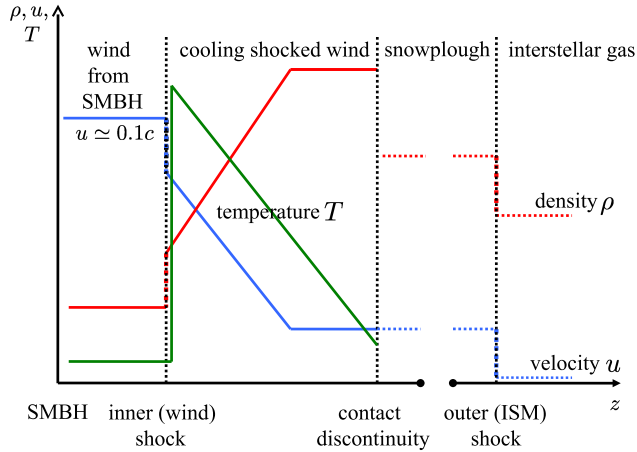} 
% \vspace*{-1.0 cm}
% \caption{Path of pre-solar grains from their stellar sources to the
%   laboratory.}
\caption{Impact of a wind from an SMBH accreting at a super--Eddington
  rate on the interstellar gas of the host galaxy: schematic view of
  the radial dependence of the gas density $\rho$, velocity $u$ and
  temperature $T$. 
%At the inner shock, the gas temperature rises
%  strongly, while the wind density and velocity respectively increase
%  (decrease) by factors $\sim 4$. Immediately outside this (adiabatic)
%  shock, the strong Compton cooling effect of the quasar radiation
%  severely reduces the temperature, and slows and compresses the wind
%  gas still further. This cooling region is very narrow compared with
%  the shock radius (see Fig. 1), and may be observable through the and
%  inverse Compton continuum and lower--excitation emission lines. The
%  shocked wind sweeps up the host ISM as a `snowplough'. This is much
%  more extended than the cooling region (cf Fig.1), and itself drives
%  an outer shock into the ambient ISM of the host. Linestyles: red,
%  solid: wind gas density $\rho$; red, dotted: ISM gas density $\rho$;
%  blue, solid, wind gas velocity $u$; blue, dotted, ISM gas velocity
%  $u$; green, solid, wind gas temperature $T$. The vertical dashed
%  lines denote the three discontinuities, inner shock, contact
%  discontinuity and outer shock.
}
   \label{fig2}
\end{center}
\end{figure}

%\begin{figure*}

%\centerline{\psfig{file=diag2.eps,width=0.5\textwidth,angle=0}}
%{epsfxsize3cm \epsfbox{diag2.eps}
%\label{}
%\end{figure*}

The required shock cooling is supplied by the inverse Compton effect
of the quasar's radiation field (King, 2003). This field
typically has Compton temperature $T_c\sim 10^7$~K, whereas the formal
temperature at the inner adiabatic shock is $m_pv^2/k \sim
10^{11}$~K. The quasar radiation cools the inner shock efficiently,
provided that this is within galaxy--scale distances from the centre
(King, 2003). Inverse Compton cooling should produce a component in
the quasar spectrum characterized by $kT_c \sim 1$~keV and with a
luminosity $\sim \mo v^2/2 \simeq 0.05\le$, i.e. about $5\%$ of the
quasar's bolometric output. Note that even if the quasar becomes
sub--Eddington, leaving a wind persisting for a lag time $10 R_{\rm
  shock}/c$, its radiation field is still able to cool the shock
efficiently.

The gas density jumps by a factor $\sim 4$ at the adiabatic
shock, accompanied by a velocity drop by the same factor. It is then
strongly compressed in the cooling region while the velocity slows to
low values (see Fig. 2). Since the cooling is efficient the whole
region is very thin compared with the shock radius $R_{\rm shock}$, and
we can regard the shock as locally plane. The Rankine--Hugoniot
relations across this isothermal shock then show that the mass flow
rate $\rho v$ remains constant, while the postshock gas pressure drops
to the value
\begin{equation}
P_{\rm ram} = \rho v^2 = {\dot Mv\over 4\pi bR_{\rm shock}^2} \simeq {\le \over
  4\pi bR_{\rm shock}^2c},
\label{ram}
\end{equation}
i.e. the preshock ram pressure.  With a constant cooling time, as
expected, the postshock temperature and velocity $u$ drop
approximately linearly with distance behind the shock, and the density
rises as $1/u$, strongly increasing its emission measure. The gas is
likely to be in photoionization equilibrium as it has low optical
depth to the quasar radiation, and the increased densities imply short
recombination times. The mass conservation equation and
ionization parameter (\ref{ion}) combine to give
\begin{equation}
{l_i u \over \xi} = {\rm constant}
\label{excit}
\end{equation}
in this region. {\it We thus expect a correlation between velocity and
  excitation.} The rapid cooling in this region implies a rapid
transition between the immediate postshock regime ($\sim v/4$, keV
excitation) and the much slower and cooler compressed state. There is
direct observational evidence for this cooling shock in NGC 4051
(Pounds et al., in prep). Pounds et al (2004) had already noted a
correlation of outflow velocity with ionization in this source.
\section{Dynamics}

Given the basic structure sketched in the last Section, we can
investigate how the shock pattern moves through the interstellar
medium of the host galaxy. The cooled postshock gas exerts the ram
pressure (\ref{ram}) on the undisturbed interstellar medium of the
galaxy, driving an outer shock into it and sweeping it up into a
relatively dense shell of increasing mass. The equation of motion of
the shell in the momentum--driven limit is
\begin{equation}
{{\rm d}\over {\rm d}t}[M(R)\dot R] + {GM(R)[M + M_{\rm tot}(R)]\over R^2} = 
4\pi\rho v^2 = {\le\over c}
\label{motion}
\end{equation}
where 
\begin{equation}
M(R) = 4\pi\int_0^R\rho_{\rm ISM} r^2 {\rm d}r
\label{m}
\end{equation}
is the swept--up interstellar gas mass, $M$ is the black hole mass,
$M_{\rm tot}= M(R)/f_g$ is the total mass within radius $R$ (including
any dark matter), and $f_g$ is the gas fraction (note that in eqn (2)
of King, 2005 the suffix `tot' was inadvertently missed off the
relevant quantity). 
%Multiplying through by $M(R)\dot R/GM$ we find the
%first integral
%\begin{equation}
%{[M(R)\dot R]^2\over 2GM} = {4\pi\over \kappa}\int M(R){\rm d}R -
%\int{M^2(R)\over R^2} {\rm d}R.
%\label{int}
%\end{equation}
%
%
%The equation of motion (\ref{motion}) takes different forms depending
%on which part of the host galaxy the shell has reached. 
%
Far from the
black hole (i.e. for $R > R_{\rm inf}$) the dark matter term $M_{\rm
  tot}$ becomes dominant in the equation of motion (\ref{motion}), and
we can drop the black hole mass term involving $M$. 
%The condition that
%the shell should just be able to escape to infinity specifies a
%relation between the black hole mass $M$ and the parameters of the
%galaxy potential, particularly the velocity dispersion, i.e. an $M -
%\sigma$ relation. For a general mass distribution $M_{\rm tot}(R)$ we
%can use the first integral (\ref{int}) to do this. However 
For a
simple isothermal potential the equation of motion has the analytic
solution
\begin{equation}
\rsh^2 = \left[{G\le\over 2f_g\sigma^2c} - 2(1-f_g)\sigma^2\right]t^2
+ 2R_0v_0t + R_0^2
\label{shell}
\end{equation}
where $R_0, v_0$ are the position and speed of the shell at time $t=0$ (King,
2005). For large times the first term dominates, and the shell can reach
arbitrarily large radii if and only if the black hole mass exceeds the
critical value
\begin{equation}
M_{\sigma} = {f_g(1-f_g)\kappa\over \pi G^2}\sigma^4 \simeq
{f_g\kappa\over \pi G^2}\sigma^4.
\label{msig2}
\end{equation}
This is very close to the observed $M - \sigma$ relation
(cf King, 2005). At sufficiently large radii the quasar radiation
field is too dilute to cool the wind shock, and the shell accelerates
beyond the escape value, cutting off the galaxy and establishing the
black--hole mass -- bulge--mass relation (cf King, 2003, 2005). 

We note that (\ref{mom}) implies a kinetic energy rate
\begin{equation}
{1\over 2}\mo v^2 \simeq {v\over c}\le \simeq {\eta\over 2}\le \simeq 0.05\le
\label{kin}
\end{equation}
implying a mechanical `energy efficiency' $\eta/2\simeq 0.05$ wrt
$\le$. Cosmological simulations typically adopt such values in order
to produce an $M-\sigma$ relation in agreement with observation
(e.g. di Matteo, 2005). This implicitly means that they adopt the
single--scattering momentum relation (\ref{mom}). We shall see below
that there must also be an implicit assumption of momentum rather than
energy driving, i.e. that the wind interacts with the host galaxy
through its ram pressure rather than its energy.

\section{Energy--Driven Outflows}

We see from the reasoning of the last Section that the interaction
beween the quasar wind and its host establishing the $M - \sigma$
relation is -- crucially -- `momentum--driven' rather than
`energy--driven'. This equivalent to requiring efficient shock
cooling. An energy--driven shock (e.g. Silk \& Rees, 1998) would
result in a much smaller black hole mass for for a given $\sigma$ than
observed. Instead of the momentum rate $\le/c$ balancing the weight of
swept--up gas $4f_g\sigma^4/G$, which is what produces the
momentum--driven relation (\ref{msig2}), an energy--driven shock would
equate the energy deposition rate to the rate of working against this
weight. In the near--Eddington regime the result is
\begin{equation}
{1\over 2}\mo v^2 \simeq {\eta\over 2}\le = 2{f_g\sigma^4\over G}.\sigma
\label{work}
\end{equation}
i.e.
\begin{equation}
M({\rm energy}) \simeq {2f_g\kappa\over \eta\pi G^2 c}\sigma^5 =
{2\sigma\over \eta c}M_{\sigma} = 3\times 10^6\msun\sigma^5_{200},
\label{energysig}
\end{equation}
which lies well below the observed relation. The coupling
adopted in cosmological simulations evidently ensures that the
interstellar medium feels the outflow momentum rather than its energy,
in addition to the `energy efficiency' $\sim \eta/2 \simeq 0.05$ noted
above.

\section{Galaxy--wide high--velocity outflows}

On large scales the outflows described in Section 5.2 above all have (outer)
shock velocities limited by the bulge velocity dispersion $\sigma$. Yet
optical and UV observations of various types of galaxies (Holt et
al., 2008; Tremonti et al., 2007)
give clear evidence of outflows with velocities of
several times this value. These cannot be the central quasar winds with $v
\sim 0.1c$ discussed in Section 3. 

There is a simple interpretation of such large--scale high--velocity
outflows. Consider a galaxy in which the SMBH has reached the mass
$M_{\sigma}$ given by eqn (\ref{msig2}), with the cosmic gas fraction
$f_g \simeq 0.16$. Its bulge gas will probably be severely
depleted. In a subsequent minor accretion event triggering AGN
activity, the effective gas fraction in the bulge will be $f'_g <
f_g$. If accretion on to the SMBH becomes super--Eddington for a time
$> 10^5$~yr, the AGN must drive an outflow shock beyond the radius
$\rinf$. However because of the discrepancy between $f_g$
(establishing the black hole mass), and $f'_g$ (the current gas
fraction), the shell radius now obeys a modified form of the analytic
solution (\ref{shell}), namely
\begin{equation}
\rsh^2 = \left[{G\le\over 2f'_g\sigma^2c} - 2(1-f'_g)\sigma^2\right]t^2
+ 2R_0v_0t + R_0^2
\label{shell2}
\end{equation}
where the $\le$ term involves $f_g$ rather than $f'_g$. Thus at large
$t$ we have
\begin{equation}
\rsh^2 = 2\left[{f_g\over f'_g}(1-f_g) - (1-f'_g)\right]\sigma^2t^2
\simeq 2{f_g\over f'_g}\sigma^2t^2 
\label{shell3}
\end{equation}
where we have taken $f'_g << f_g < 1$ in the last form. This shows
that the shell reaches velocities
\begin{equation}
\simeq
(2f_g/f'_g)^{1/2}\sigma > \sigma, 
\label{vel}
\end{equation}
because its inertia is much lower than the one previously expelled by
the Eddington thrust in the accretion episode which defined the SMBH
mass. If at some point the AGN activity turns off, we can match
another solution of the form (\ref{shell}), but with $\le$ formally $=
0$, to the solution (\ref{shell2}). This solution reveals that a
coasting shell stalls only at distances $\sim (f_g/f'_g)^{1/2}$ times
its radius $R_0$ at the point when AGN activity ceased, and thus
persists for a timescale $R_0/\sigma \sim 10^8$~yr.

Episodic minor accretion events of this type therefore naturally produce
large--scale outflows with velocities $> \sigma$. Moreover, since they
persist as fossil winds long after the AGN has become faint, they can
have total momentum considerably higher than could be driven by the
{\it current} AGN radiation pressure, i.e. well in excess of the
apparent momentum limit. A recent paper (King, 2009) gives more
details of the expected outflows.

\section{Conclusion}

AGN accretion appears to involve a series of very small--scale,
chaotic events, whose gas flows have no correlation with the
large--scale structure of the galaxy or with each other. The accreting
gas has extremely low specific angular momentum and is presumably only
a small fraction of the gas involved in a galaxy merger.

The growth of SMBH through accretion requires Eddington accretion
episodes in AGN to be common. Mass and momentum conservation then
imply winds with velocities $v \sim 0.1c$ and the presence of
resonance lines of helium-- and hydrogenlike iron.  The wind shocks
and cools as it interacts with the interstellar medium of the host
galaxy, and may be observable in an inverse Compton continuum and
lower--excitation emission lines with lower velocities.  The shocked
wind begins to sweep up the galaxy ISM once the black hole mass
reaches the value $M_{\sigma}$ implied by the $M - \sigma$ relation,
preventing growth beyond this mass. If the shocked gas did not cool as
stated above, the resulting (`energy--driven') outflow would imply a
far smaller SMBH mass than actually observed. Minor accretion events
with small gas fractions can produce galaxy--wide outflows, including
fossil outflows in galaxies where there is little current AGN
activity.

\section{Acknowledgments}

I thank Ken Pounds and Sergei Nayakshin for illuminating discussions
and the Royal Society for a travel grant.

\end{document}